          [2001/04/25 1.1 (PWD)]
\documentclass[a4paper,twocolumn]{esapub} 

\usepackage{times,graphicx}

\title{Hard X--ray emission from Serpens X-1 as observed by
{\it INTEGRAL}}

\author{N. Masetti$^{1}$}
\author{L. Foschini$^{1}$}
\author{E. Palazzi$^{1}$}
\author{L. Amati$^{1}$}
\author{E. Caroli$^{1}$}
\author{G. Di Cocco$^{1}$}
\author{F. Frontera$^{1,2}$}
\author{M. Orlandini$^{1}$ \\
{\it on behalf of a larger collaboration}}

\affil{$^{1}$Istituto di Astrofisica Spaziale e Fisica Cosmica (IASF) -- 
Sezione di Bologna, CNR, Italy \\
$^{2}$Dipartimento di Fisica, Universit\`a di Ferrara, Italy}

\def\fs{\hbox{$\,.\!\!^{\rm s}$}}

\begin{document}


\maketitle

\begin{abstract}

We here report results of an {\it INTEGRAL} observation of the X--ray
burst and atoll source Ser X-1 performed on May 2003. The object was
observed for a total of 400 ks but nearly 8$^\circ$ off-axis due to the
amalgamation with an observation of SS 433, the pointing target source.
Ser X-1 was detected up to 30 keV with unprecedented positional accuracy
for a high-energy emission; a sharp spectral drop is evident beyond this
energy. Significant variability is seen in the 20--30 keV light curve.  
Comparison with previous observations indicates that the source was in its
high (banana) state and displayed a soft spectrum during the {\it
INTEGRAL} pointing. A (non simultaneous) broadband radio-to-$\gamma$--rays
broad-band spectral energy distribution for Ser X-1 is also presented for
the first time.

\end{abstract}

\begin{figure*}
\centering

\includegraphics[width=16cm]{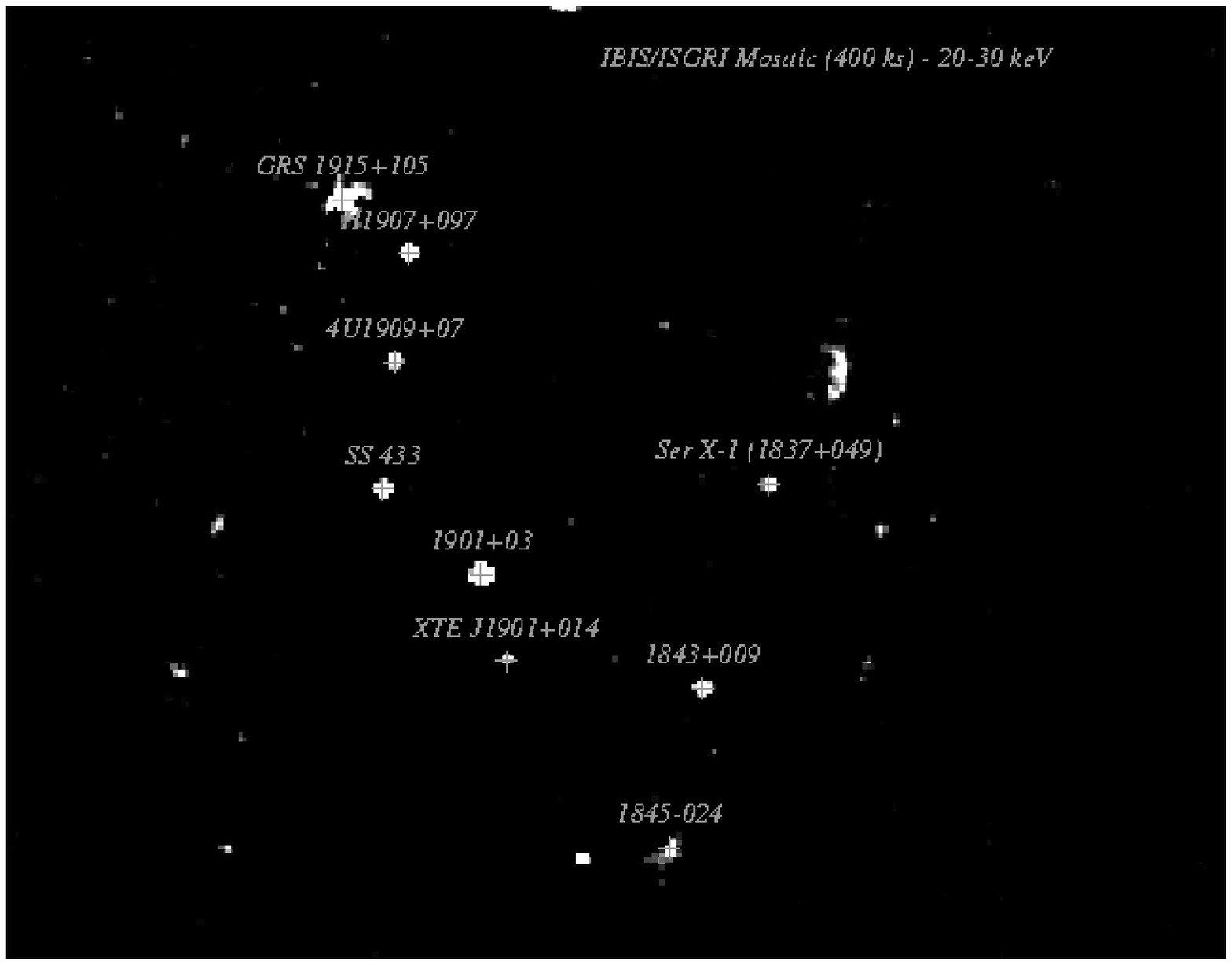}

\includegraphics[width=16cm]{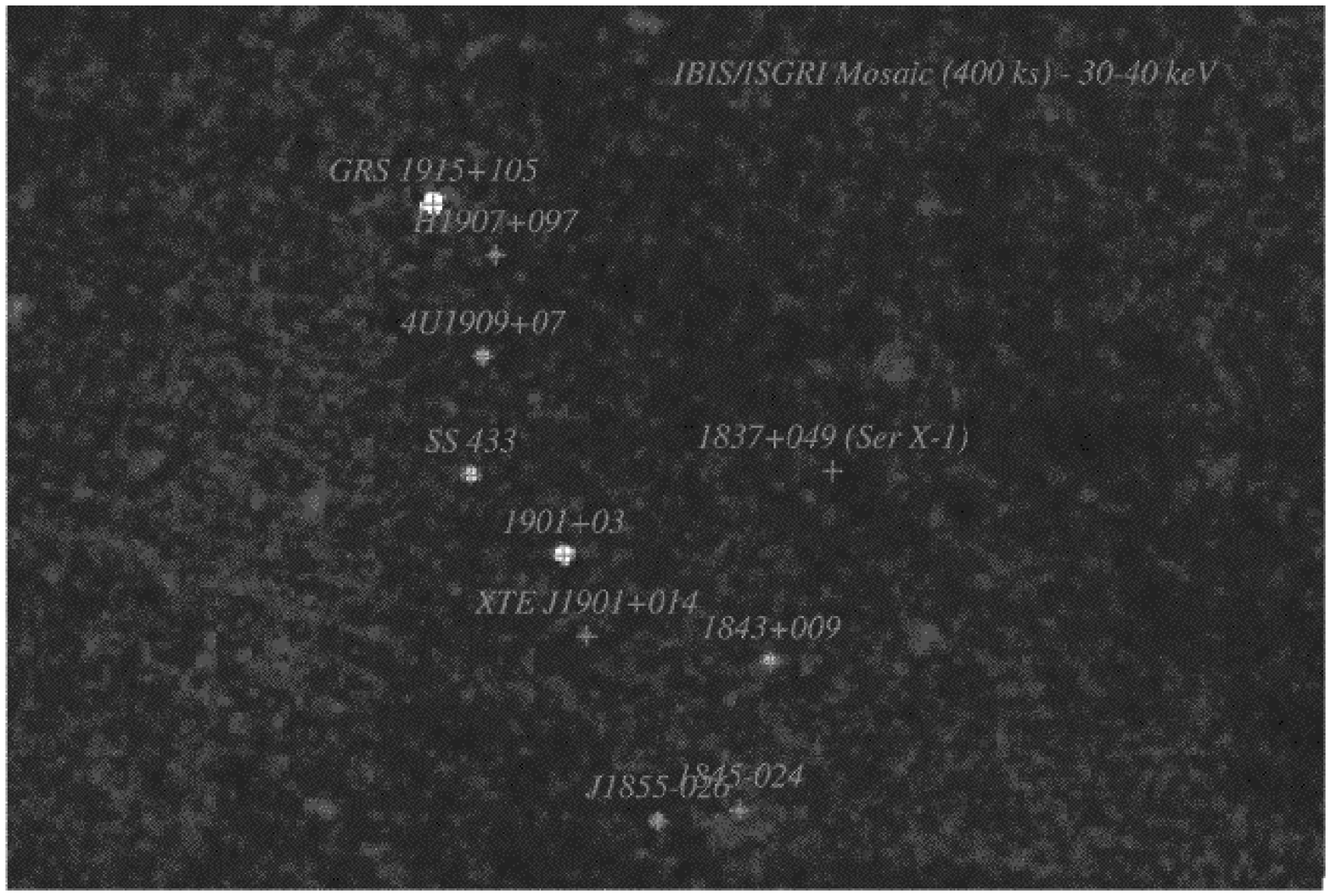}

\vspace{-.3cm}
\caption{Part of the mosaic image of the Ser X-1 field as observed by 
ISGRI in the 20--30 keV band {\rm (upper panel)} and in the 30--40 keV band 
{\rm (lower panel)} and accumulated over the whole {\it INTEGRAL} pointing 
(400 ks on-source time). The field is about 22$^{\circ}$$\times$28$^{\circ}$. 
North is at top, East to the left. Ser X-1 is detected at $\sim$25$\sigma$ 
significance in the 20--30 keV range, while no detection is apparent in 
the 30--40 keV band.}
\end{figure*}

\section{Introduction}

The low-mass X--ray binary Ser X-1, or 4U 1837+04 is known to host a 
neutron star as the accreting object; it is classified as an
Atoll source (e.g., Liu et al. 2001). Archival {\it EXOSAT} data (Seon \&
Min 2002) showed this object in the banana (i.e. high intensity) state
during the observations. More recent {\it BeppoSAX} and {\it RXTE}
pointing (Oosterbroek et al. 2001) caught the source while it was again in
a high activity state, with an unabsorbed flux (1--200 keV) of
8.0$\times$10$^{-9}$ erg cm$^{-2}$ s$^{-1}$. Its spectrum was 
well described by a combination of a blackbody-like and Comptonization
(Titarchuk 1994) models to account for the observed hard tail. A
reflection component could not be excluded, but the data quality could not
provide a definitive conclusion. Up to now, however, the X--ray emission
observed from Ser X-1 with the most recent high-energy missions was never
well representative of the hard (island) state which is generally seen
when Atoll sources are undergoing the low intensity phase (Barret 2001).

Several X--ray bursts (lasting tens of seconds at most) were also
detected from Ser X-1; moreover, during 2001, {\it BeppoSAX} 
pinpointed a very long ($\sim$4 hours) X--ray burst (Cornelisse et al.
2002), making this source join the group of `superbursters' (see Kuulkers 
2003 for a review). By studying X--ray bursts observed with {\it Einstein},
Christian \& Swank (1997) deduced a distance to the source of 8.4 kpc.
This implies a 1--200 keV luminosity of 6.7$\times$10$^{37}$ erg 
s$^{-1}$ during the {\it BeppoSAX} observation of Oosterbroek et al. 
(2001), which means roughly one third of the Eddington luminosity for a 
neutron star.

The optical counterpart to Ser X-1, located in a crowded stellar field
(Thorstensen et al. 1980), was correctly identified by Wachter (1997) and,
subsequently, spectroscopically confirmed and studied by Hynes et al.
(2004). Very recently, the radio counterpart was discovered with the VLA
(Migliari et al. 2004).

We here report on a observation of Ser X-1 performed with the
INTErnational Gamma--RAy Laboratory ({\it INTEGRAL}; Winkler et al. 2003)  
on May 2003, i.e. less than 7 months after the launch of this spacecraft.  
The high spectral sensitivity of the high-energy instruments onboard this
satellite are optimal to study the behaviour of the hard X--ray tail of
this source in case of its presence. A more complete analysis of these
data can be found in Masetti et al. (2004).

\section{Observations and analysis}

The observations presented here were acquired during Revolutions 67 to 69,
i.e. between 12:00 UT of 3 May 2003 and 09:26 UT of 9 May 2003, for a
total on-source time of of 400 ks, with the SPI (Vedrenne et al. 2003) and
IBIS (Ubertini et al. 2003) instruments onboard {\it INTEGRAL}. These
detectors allow an actual 20 keV -- 10 MeV spectral coverage altogether.
In particular, the IBIS telescope is composed of two detector layers:
ISGRI (Lebrun et al. 2003), optimized for the energy range 20--200 keV,
and PICsIT (Di Cocco et al. 2003; Labanti et al. 2003), covering the range
from 175 keV to 10 MeV. IBIS is a coded-mask imager with a wide field of
view 29$^{\circ}$$\times$29$^{\circ}$ at zero response;  
9$^{\circ}$$\times$9$^{\circ}$ fully coded) and a high angular resolution
of 12$'$, sampled in 5$'$/pixel in ISGRI and 10$'$/pixel in PICsIT. In
this preliminary analysis we did not consider the data from JEM-X (Lund et
al. 2003) and SPI (Vedrenne et al. 2003), the other two high-energy
instruments onboard {\it INTEGRAL}.

Data were acquired with a spacecraft rectangular 5$\times$5 dithering
pattern mode. Unfortunately, due to pointing constraints -- the present
observation was amalgamated with a pointing centered on the source SS 433
(see Cherepashchuk et al. 2003) located nearly 8$^{\circ}$ away --, Ser
X-1 was outside the field of view of the fourth instrument carried by
{\it INTEGRAL}, i.e. the optical camera OMC (Mas-Hesse et al. 2003), and
therefore could not be observed by it.

The IBIS data reduction and analysis was performed with the Offline
Scientific Analysis\footnote{Available through the {\it INTEGRAL} Science 
Data Centre at: \\ {\tt
http://isdc.unige.ch/index.cgi?Soft+download}} (OSA), version 3.0, whose
algorithms are described in Goldwurm et al. (2003). The count rates
extracted from the standard pipeline were normalized to those measured
during the observation of calibration of the Crab, performed from 22 to 24
February 2003, in order to obtain a flux evaluation in term of physical
units.

A further correction, as described in Goldoni et al. (2003), has been 
applied to take into account effects introduced by the off-axis position 
of the source with respect to the detector field of view.

\section{Results}

\subsection{IBIS detection}

Ser X-1 was well detected by ISGRI in the 20--30 keV band at a
$\sim$25$\sigma$ level in the whole 400 ks observation. the source shows 
an average count rate of 0.84 cts s$^{-1}$ in this band, corresponding to 
a flux of 6.2$\times$10$^{-11}$ erg cm$^{-2}$ s$^{-1}$.

No detection of Ser X-1 was instead obtained in the 30--40 keV (ISGRI) and
220--280 keV (PICsIT) ranges down to 3$\sigma$ upper limits of
6$\times$10$^{-12}$ erg cm$^{-2}$ s$^{-1}$ and 1$\times$10$^{-10}$ erg
cm$^{-2}$ s$^{-1}$, respectively. The upper panel of Fig. 1 shows the
image of the Ser X-1 field as obtained by ISGRI in the 20--30 keV during
our {\it INTEGRAL} observation, while the lower panel of Fig. 1 reports
the field of Ser X-1, again imaged by ISGRI, in the 30--40 keV band.

The ISGRI source detection was at coordinates (J2000) $\alpha$ = 
18$^{\rm h}$ 40$^{\rm m}$ 00$\fs$0, $\delta$ = +05$^{\circ}$ 02$'$ 06$''$. 
The significance of the detection implies a 90\% confidence positional
accuracy of 1$'$ (Gros et al. 2003). The position is fully consistent with
that of the optical counterpart MM Ser (Thorstensen et al. 1980; Wachter
1997) as well as with the radio position (Migliari et al. 2004).

\begin{figure}[t!]
\vspace{-0.7cm}
\includegraphics[width=8.5cm]{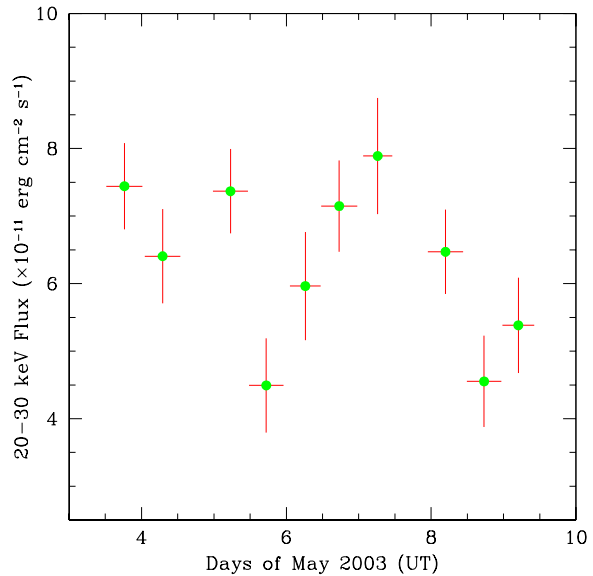}
\vspace{-0.7cm}
\caption{20--30 keV light curve of Ser X-1, rebinned at $\sim$40 ks, as 
observed by ISGRI onboard {\it INTEGRAL} during the pointing presented 
here.}
\end{figure}

\begin{figure}[t!]
\hspace{-0.5cm}
\includegraphics[width=8.5cm]{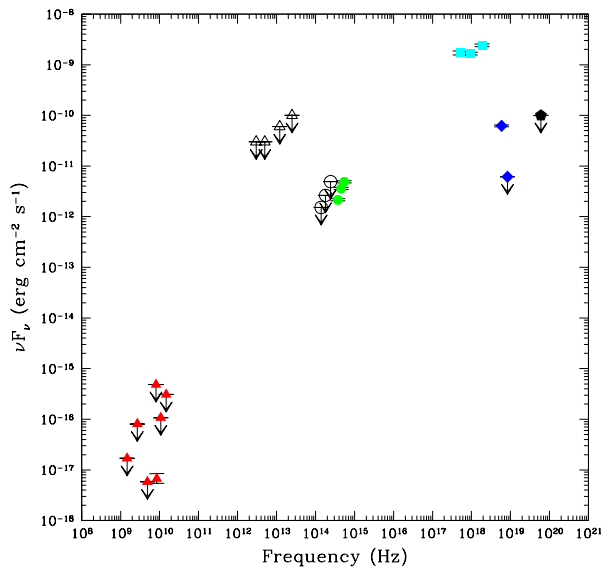}
\vspace{-1.3cm}
\caption{Broadband SED of Ser X-1 constructed with simultaneous data
and upper limits from ISGRI (filled diamonds) and PICsIT (filled pentagon)
plus ASM data (filled squares), together with non-simultaneous
optical measurements (filled dots), near-infrared and far-infrared upper 
limits (open circles and open triangles, respectively) and radio data 
(filled triangles). Error bars indicate the 1$\sigma$ uncertainty for each 
measurement.
}
\end{figure}

\subsection{ISGRI 20--30 keV light curve} 

The ISGRI light curve in Fig. 2 was obtained by dividing the whole
observation in 10 intervals of about 40 ks each.  We chose this temporal
resolution as it turned out to be the best tradeoff between the time
sampling and the S/N ratio of each bin. Significant (99.99\% confidence
level) fluctuations are present in the light curve. This behaviour
confirms the variability findings of Oosterbroek et al. (2001) in the
2--10 keV as observed with {\it BeppoSAX}.

\subsection{Spectral Energy Distribution (SED)}

By using the ISGRI and PICsIT results, {\it RXTE}/ASM data obtained
simultaneously with the {\it INTEGRAL} observations, and (non
simultaneous) archival optical data (Wachter 1997), 2MASS near-infrared
(Skrutskie et al. 1997), IRAS far-infrared (Beichman et al. 1988), and
radio upper limits (Migliari et al. 2004; Wendker 1995 and references
therein) we constructed, for the first time for Ser X-1, a broadband SED
(Fig. 3) spanning from radio to $\gamma$--ray frequencies. 
Although non-simultaneous with the high-energy part of the spectrum, these 
lower frequency data and upper limits can give us a general description of 
the overall SED of this source thanks to the relative stability of its
emission with time.

Optical and near-infrared data were corrected assuming a color excess 
$E(B-V)$ = 0.8 (e.g. Hynes et al. 2004 and references therein) and 
converted into fluxes using the normalizations by Fukugita et al. (1995) 
for the optical and the ones referring to the 2MASS\footnote{These 
normalizations are available at: \\
{\tt http://www.ipac.caltech.edu/2mass/releases/ \\ /allsky/faq.html}}
for the near-infrared. Given the resolution of the 2MASS survey (pixel 
size: 2$''$) and the crowding of the Ser X-1 field (e.g. Wachter 1997), we 
considered the 2MASS near-infrared detections as conservative upper limits 
to the actual source fluxes in the $JHK$ bands.

\section*{Acknowledgements}

The results presented in this paper are based on observations with {\it 
INTEGRAL}, an ESA project with instruments and science data centre funded 
by ESA member states (especially the PI countries: Denmark, France, 
Germany, Italy, Switzerland, Spain), Czech Republic and Poland, and with 
the participation of Russia and the USA.
We thank Simone Migliari for having communicated us the Ser X-1 radio
detection result prior to publication. Pavel Binko is acknowledged for
the help in the {\it INTEGRAL} data retrieval from the ISDC archive.  
This work has made use of the NASA Astrophysics Data System Abstract
Service, of the SIMBAD database, operated at CDS, Strasbourg, France,
and of data products from the 2MASS.
ASM data were provided by the {\it RXTE} ASM teams at MIT and at the
{\it RXTE} SOF and GOF at NASA's GSFC. This research was partially funded
by ASI.

\end{document}